\documentclass[aps,prl,twocolumn]{revtex4}
\usepackage{amssymb,graphicx}
\newcommand{\nn}{\nonumber}

\newcommand{\fig}[2]{\includegraphics[width=#1]{./#2}}
\newcommand{\Fig}[1]{\includegraphics[width=\columnwidth]{./#1}}
\newlength{\bilderlength}

\begin{document}
\bibliographystyle{KAY}
\title{
Statistics of static avalanches in a random pinning landscape}

\author{
Pierre Le Doussal$^{1}$, A.\ Alan Middleton$^2$, Kay J\"org Wiese$^1$ \vspace*{3mm} }

\affiliation{ $^1$CNRS-Laboratoire de Physique Th{\'e}orique de l'Ecole
Normale Sup{\'e}rieure, 24 rue Lhomond, 75005 Paris, France. \\
$^2$Physics Department, Syracuse University, Syracuse, NY
13244, USA.}

\date{\small March 6, 2008}

\begin{abstract}
We study the minimum-energy configuration of a $d$-dimensional elastic interface in a random potential tied to a
harmonic spring. As  a function of the spring position, the center of mass of the interface changes in discrete
jumps, also called shocks or ``static avalanches''. We obtain analytically the distribution of avalanche sizes
and its cumulants within an $\epsilon=4-d$ expansion from a tree and 1-loop resummation,  using functional
renormalization. This is compared with exact numerical minimizations of interface energies for  random field
disorder in $d=2,3$. Connections to the Burgers equation and to dynamic avalanches are discussed.
\end{abstract}

\maketitle

In numerous systems, the equilibrium or non-equilibrium response to perturbations is not smooth and involves
jumps, avalanches or bursts. In systems at the brink of instability, with many metastable states, it is often
self-organized and critical with power-law tails for the probability of large events. This is observed
ubiquitously in systems with heterogeneities, such as Barkhausen noise and hysteresis in magnets, field response
of superconductors, contact line of fluids, cracks, granular matter, dry friction and earthquakes. Sandpile
automata \cite{sandpilereviews} and elastic media pinned by quenched disorder \cite{fisher_review98} have been
studied as simple models for these phenomena. Relations between sandpiles and periodic interface depinning  \cite{NarayanMiddleton1994,Alava2002} and between sandpile models and loop-erased walks \cite{MajumdarDhar1992,dharLERW,priezzhev99}
have proved fruitful, especially in $d=2$, where conformal
field theory can be used \cite{cft}. Despite much effort it has proven difficult to obtain analytical results,
e.g.\ for the distribution of the size $s$ of avalanches (defined below), except in mean-field models for
sandpiles \cite{meanfield} and for random field Ising magnets \cite{dahmensethna96} as well as for a toy model
for avalanches at depinning \cite{fisher_review98}, which all yield $P(s) \sim s^{-\tau}$ with $\tau=3/2$.
Scaling arguments for sandpiles \cite{sandpilereviews,dharLERW} and for depinning
\cite{ZapperiCizeauDurinStanley1998}, were developed together with numerical analysis
\cite{MiddletonFisher1993,NarayanMiddleton1994,lubeck97}. The Functional Renormalization Group (FRG) theory for pinned systems has
led to detailed predictions for e.g.\ the roughness of interfaces but, until now, has failed to describe
discontinuous jump processes \cite{DSFisher1986,depinning,ChauveLeDoussalWiese2000a}. Hence it remains an
outstanding issue to find  a limit where mean field theory is valid, prove this, and to develop a controlled
field-theoretic expansion around it. It should allow to clarify the differences between equilibrium and
non-equilibrium avalanches, recently questioned in a model for magnetic hysteresis \cite{dahmen06}.

The aim of this Letter is to provide a first analytical calculation of the distribution of avalanche sizes in a
static, equilibrium setting, using FRG, and to compare with numerical
calculations
. It opens the way to a closely related
calculation for depinning \cite{us_dynamics}. As demonstrated in our previous work
\cite{MiddletonLeDoussalWiese2006}, a model which allows a precise FRG treatment and comparison with numerics,
both in statics and dynamics, consists of an elastic interface in a random potential, parameterized by a
(scalar) height field $u(x)$, and submitted to an external parabolic well, i.e.\ a spring, centered at $u=w$,
\begin{equation}
{\cal H}[u;w] =  \int d^d x\, \frac{1}{2}[\nabla u(x)]^2 + V(x,u(x)) + \frac{m^2}{2} [u(x)-w]^2\ .
\end{equation}
We are interested in energy minimization as $w$ is varied in a given realization of the random potential
$V(x,u)$. We denote $\hat V(w) = \min_{\{u(x)\}} {\cal H}[u;w]$ the optimal energy and $u(x;w)$ the optimal interface
position. The force per unit volume exerted by the spring is $\hat V'(w) = m^2 [w- u(w)]$, where $u(w):=L^{-d}
\int d^d x\, u(x;w)$ is the center-of-mass position and $L^d$ the volume  of the system. We study the three
important universality classes for disorder, with short range (random bond, RB), long range (random field, RF)
or periodic disorder correlator (random periodic, RP). Their definitions can be found in
\cite{MiddletonLeDoussalWiese2006} and in standard papers on pinning \cite{pinning} and FRG
\cite{DSFisher1986,depinning,ChauveLeDoussalWiese2000a}.

Although we often use the language of dynamics, one should emphasize the difference between the static problem
studied here, where the interface finds the global energy minimum for each $w$, and the dynamic one, where
$w(t)$ grows very slowly, and the interface visits a deterministic sequence of metastable states. In the scaling
limit $m \to 0$, on which we focus here, the first case is about interface configurations of zero-temperature
equilibrium, studied in \cite{MiddletonLeDoussalWiese2006}, whereas the latter one is about critical depinning,
studied in \cite{LeDoussalWiese2006a,RossoLeDoussalWiese2006a}. In the statics $u(w)$ is a (single-valued)
function, while it shows some history dependence at depinning. Despite these differences, depinning and statics
are close cousins and some differences within the FRG are found only beyond one loop
\cite{ChauveLeDoussalWiese2000a}.

As shown previously \cite{DSFisher1986,pinning,Middleton1995,NohRieger2001,MiddletonLeDoussalWiese2006}, the
optimal interface is statistically self-affine with $\overline{(u(x)-u(0))^2} \sim |x|^{2\zeta}$ and a roughness
exponent $\zeta$ which depends on the class of disorder, and with a $\epsilon=4-d$ expansion
\cite{ChauveLeDoussalWiese2000a}:  $\zeta=\epsilon/3$ for RF, $\zeta=0$ for RP, and $\zeta= 0.2083 \epsilon+
0.006 86 \epsilon^2$ for RB (and $\zeta=2/3$ in $d=1$). This holds for scales $L_c<L<L_m$, where $L_c$ is the
Larkin length (here of the order of the microscopic cutoff) and $L_m \sim 1/m$, the large scale cutoff induced
by the harmonic well. It is useful to picture the interface as a collection of $(L/L_m)^d$ regions pinned almost
independently.

We found that $u(x;w)$ is an increasing function of $w$ which can be decomposed into smooth parts, negligible in
the scaling limit $m\to 0$ and jumps (alias ``shocks'' or ``static avalanches'') as $u_x(w) = \sum_i S_i^x
\theta(w-w_i)$, where $S_i=\int d^d x\, S_i^x$ is the size of the shock or avalanche labelled $i$, and
$\theta(x)$ the unit-step function. The avalanche-size distribution, defined from an average over samples,
\begin{equation}
 \overline{ \sum_i \delta(S-S_i) \delta(w-w_i) } = \rho(S) = \rho_0 P(S)\ ,
\end{equation}
can equivalently be defined from a translational average in a given sample. Here $P(S)$ is the normalized size
distribution and $\rho_0 dw$ the average number of
shocks in an interval $dw$. The scaling ansatz
\begin{equation}
 \rho(S)=L^d m^\rho S^{- \tau} \tilde \rho(S m^{d+\zeta})
\end{equation}
is shown below to hold within the $\epsilon$ expansion and verified by our numerics. The constraint
$\overline{u'(v)}=1$ relates the shock rate to the first moment $L^{-d} \rho_0 \langle S \rangle  = 1$, where
$\langle S^n \rangle:= \int dS\, S^n P(S)$ denotes the normalized moments. It implies {for $\tau < 2$} the
exponent relation $\rho = (2-\tau) (d+\zeta)$. The distribution is qualitatively different for (i) $\tau<1$ when
a unique scale $S_m \sim m^{-(d+\zeta)}$ exists, i.e. $p(S)=S_m^{-1} \tilde p(S/S_m)$, and (ii) $1<\tau<2$,
where
\begin{equation}
 P(S) = C_\tau S_0^{-1} (S/S_0)^{- \tau} f(S/S_m)\ ,
\end{equation}
and typical avalanches are of the order of the microscopic (UV) cutoff $S_0$, while moments $\left<S^p\right>$
with $p>\tau-1$ are controlled by rare avalanches of size $\sim S_m$, the large-scale cutoff.

As discussed recently \cite{LeDoussal2006b}, shocks for manifolds are a natural generalization of shocks in
decaying Burgers turbulence, seen as their $d=0$ limit (with, for the RF case, $\tau=1/2$ in that limit). The
force field $\hat V'(w) = m^2 [w- u(w)]$ generalizes the Burgers velocity field. Its $n$-point connected
correlation $\overline{\hat V'(w_1)\ldots \hat V'(w_n)}^c= L^{-(n-1) d} (-1)^n \hat C^{(n)}(w_1,\ldots,w_n)$
obeys FRG equations, which generalize the hierarchies of Burgers multi-point correlations. Shock-size moments
can be extracted from their non-analytic part: E.g.\ the second moment is contained in the cusp of the disorder
correlator $\Delta(w)=\hat C^{(2)}(w,0)$ of the FRG, as seen from the relation $- m^{-4} L^{-d} \Delta''(w) =
\overline{u'(w) u'(0)}=L^{-2 d} \overline{\sum_i S_i^2 \delta(w-w_i)} \delta(w) + \text{smooth function}$, which
upon integration gives $- 2 \Delta'(0^+) = m^4 \langle S^2 \rangle/\langle S \rangle$. This generalizes to
higher cumulants \cite{LeDoussal2006b} $K^{(n)}(w):=m^{2n} L^{(n-1)d} \overline{[u(w)-w-u(0)]^n}^c$ (the linear
cusp of $K^{(3)}$, measured in \cite{MiddletonLeDoussalWiese2006}, generalizes Kolmogorov's law). Hence we can
compute the avalanche-size distribution using the generating function
\begin{eqnarray}
&&
L^{-d} (\overline{e^{ \lambda L^d  [u(w)-w-u(0)]}} - 1 ) =  Z(\lambda) w + O(w^2) \nonumber\\
&& Z(\lambda) = \frac{1}{\langle S \rangle} \left(\langle e^{\lambda S} \rangle -1 - \lambda \langle S \rangle \right)\qquad
\end{eqnarray}
for $w>0$. It assumes a linear cusp (i.e.\ a finite density of shocks) and generalizes to depinning
\cite{us_dynamics}.

We have computed the leading non-analyticity of the functions  $\hat C^{(n)}$ and $K_n(w)$  \footnote{Cumulants and
moments have the same non-analyticity.} from (i) a Legendre transform of the replicated effective action
$\Gamma$ computed order by order in $\epsilon$;  (ii) a direct perturbative expansion without replica. The
calculation is more involved than usually for FRG: the size distribution already at order $O(\epsilon^0)$ requires a summation of all tree diagrams. The latter could
be termed mean field, but with the proviso that the scale of $S$ involves $\Delta'(0^+)$ computed to
$O(\epsilon)$.
Here we   compute  to order
$O(\epsilon)$, which amounts to sum all trees and single loops; for details see \cite{kwpld}. The main result is that $Z(\lambda)$ satisfies a
remarkable self-consistent equation to one loop
\begin{equation} \label{selfcons}
\tilde Z (\lambda) = \lambda + \tilde Z (\lambda)^{2}  + \alpha \sum_{n \geq 3} (n+1) 2^{n-2} i_{n} \tilde Z
(\lambda)^{n}\ ,
\end{equation}
where $Z(\lambda) = \frac{m^4}{|\Delta' (0^{+})|} \tilde Z (\lambda m^{-4}|\Delta' (0^{+})|) -
\lambda$, $i_n=I_n/(\epsilon \tilde I_2)$, $\tilde I_n=\int_k(k^2+1)^{-n}$, $\alpha=-\epsilon \tilde I_2
m^{-\epsilon} \Delta''(0^{+})$. It can graphically be written as
$$
\!\!\fig{8cm}{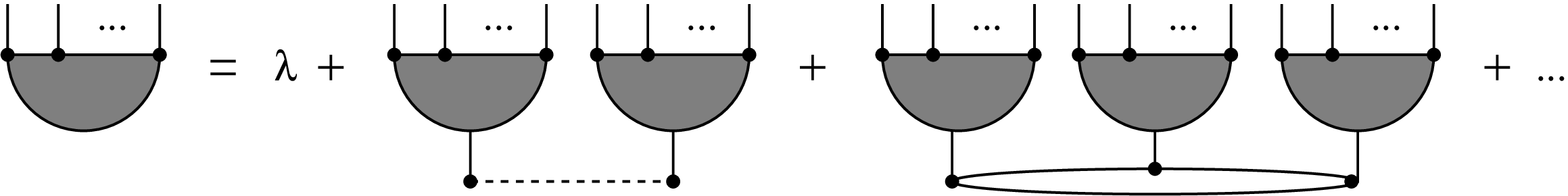}\ .
$$%
\begin{figure}[tbp]
\begin{center}
\Fig{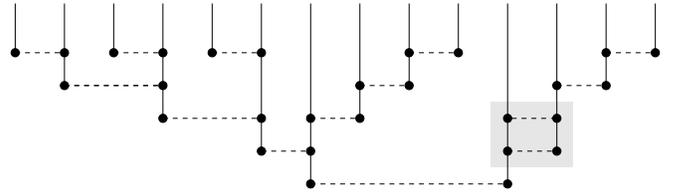}
\caption{Example of a diagram at MF level, as generated by Eq.~(\ref{selfcons}) at $\alpha=0$. It contains a correction to disorder i.e.\ $\Delta'(0^{+})$ at 1 loop (shaded in gray).}
\label{f:resum}
\end{center}
\end{figure}%
The type of
resummed diagrams is presented on figure \ref{f:resum}.
Since $\alpha=O(\epsilon)$, to leading order one solves (\ref{selfcons}) setting $\alpha=0$.  This yields $\tilde Z(\lambda)=\frac{1}{2}(1-\sqrt{1-4
\lambda})$, identical to the generating function of the number of rooted binary planar trees with $n$ leaves
\footnote{Equals the number of rooted planar trees with $n+1$ bonds and arbitrary coordination, the 
Catalan numbers.}, and a size distribution, with $\tau=3/2$:
\begin{equation} \label{meanfield}
P(S) = \frac{\langle S \rangle}{2 \sqrt{\pi}}  S_m^{- 1/2} S^{- 3/2} e^{- S/(4 S_m)}\ .
\end{equation}%
\begin{figure}[t]
\Fig{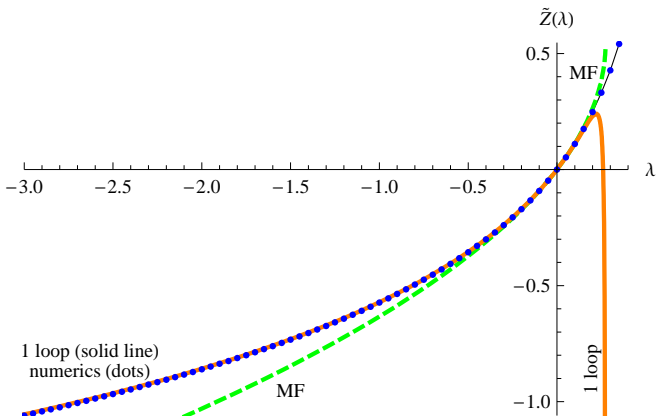} \caption{$\tilde Z(\lambda)$ for RF, $d=2$. The mean field and one loop analytical curves
are given in the text, $\lambda$ being rescaled so as to reproduce the numerically measured second moment (given
by the curvature of $\tilde Z(\lambda)$). While the MF result differs substantially from the numerical
measurement, the 1-loop curve, i.e. (\ref{a61}) with $\alpha$ given by (\ref{alpha}) setting $\epsilon=2$ and
$\zeta_{1}=1/3$, coincides for all negative $\lambda$, and almost up to the singularity for $\lambda>0$.
Changing $\epsilon$ from $2$ to $2.1$ or $1.9$ already results in a visible disagreement for $\lambda<0$.}
\end{figure}%
This is valid for $S \gg S_{0}$, such that \footnote{Note that for $\lambda \to - \infty$,  $Z(\lambda) \sim_{} -
S_m^{\tau-2} |\lambda|^{\tau-1}$ in the scaling regime and should converge to $-1/\langle S \rangle \sim -
S_0^{1-\tau} S_m^{\tau-2}$ for $\lambda \sim 1/S_0$ in the UV cutoff regime.} the moments with $p>\frac{1}{2}$
satisfy $\langle S^p \rangle/\langle S \rangle=a_p S_m^{p-1}$ with $a_p=2^{2p-2} \pi^{-1/2}
\Gamma(p-\frac{1}{2})$, independent of the non-universal small-scale cutoff $S_0$. Hence the rigorous summation
of tree diagrams in the FRG yields the same $P(S)$ as that of  a mean-field toy model for dynamic avalanches
\cite{fisher_review98} and that of mean-field sandpiles \cite{meanfield}. In addition, since the FRG is a first
principle method, it predicts $S_m= c m^{-d-\zeta}$ where $c=(\epsilon \tilde I_2) |\tilde \Delta'(0^{+})|$ is
obtained from the FRG fixed point for the rescaled correlator $\tilde \Delta(u)=(\epsilon \tilde I_2)
m^{-\epsilon+2\zeta} \Delta(u m^{-\zeta})$ and depends on the universality class \cite{kwpld}. Since $\tau>1$
the scale $\langle S \rangle \sim S_0^{\tau-1} S_m^{2-\tau}$ remains undetermined and UV cutoff dependent.
Eq.~(\ref{selfcons}), seen as a convolution equation for $P(S)$, may allow to put the physical picture in
\cite{fisher_review98} on a more rigorous footing.

To next order in $\epsilon$ we solve Eq.~(\ref{selfcons}) which includes higher branchings with a universal
dimensionless rate
\begin{equation}\label{alpha}
\alpha= - \frac{1}{3} (1- \zeta_1) \epsilon
\end{equation}
at the fixed point, where $\zeta=\zeta_1 \epsilon + O(\epsilon^2)$ and $i_n=1/2(n-1)(n-2)$ in $d=4$. It yields
\begin{eqnarray}\label{a61}
\!\!\tilde Z (\lambda) \!\!&=&\!\! \frac{1}{2} \left[1-\sqrt{1-4 \lambda }\right]+\frac{\alpha}{4
\sqrt{1-4 \lambda }} \Big[\log (1-4 \lambda ) \times\nonumber\\
&& \quad\times (3 \lambda +\sqrt{1-4
\lambda }-1) -2(2 \lambda +\sqrt{1-4 \lambda }-1)\Big] \nn\\
&&+O\left(\alpha ^2\right) \ ,
\end{eqnarray}
from which one can calculate the universal ratios:
\begin{figure}[t]
\begin{center}
{\Fig{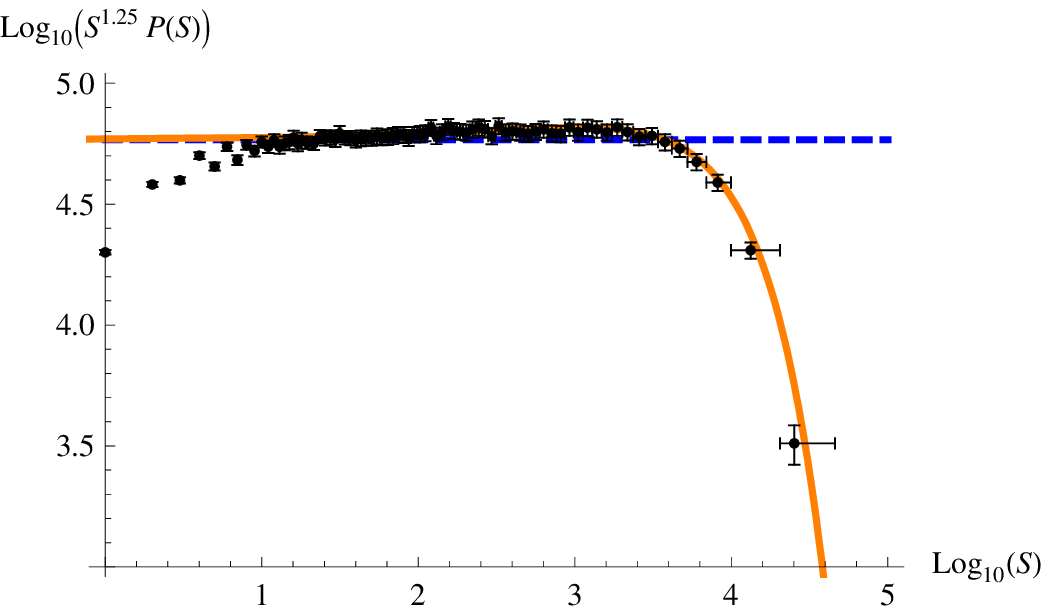}\hspace{-7.5cm}\raisebox{20mm}[0mm][0mm]{\parbox{0mm}{\fig{5cm}{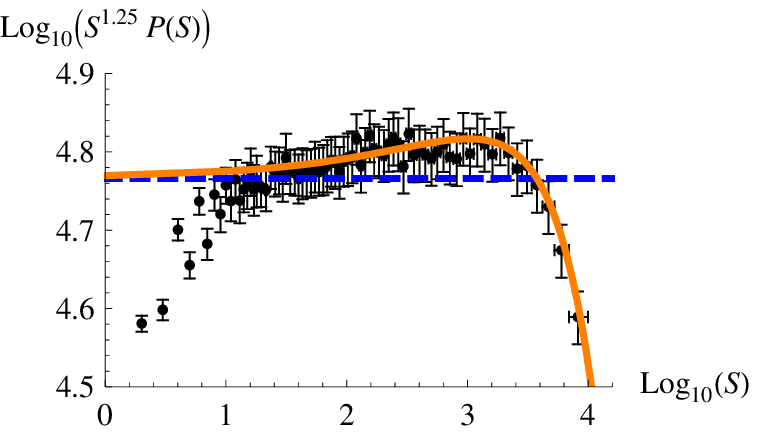}}}\hspace{7.5cm}}
\caption{Numerically computed normalized avalanche distribution $P(S)$, for random-field disorder and $d=2$
($\zeta=2/3$), multiplied by $S^{\tau}$ with $\tau=1.25$ (from (\ref{conj})) to emphasize deviations from the
 power law $P(S)\sim S^{-1.25}$. Error-bars are $3\sigma$-errors for $P(S)$ and the size of the box for
$S$. The solid curve is a 1-parameter fit to Eqn.~(\ref{final}), with $S_m=3500$, $\tau=1.25$, $\alpha$ given by
(\ref{alpha}) with $\zeta_1=1/3, \epsilon=2$ and the corresponding values for $B,C$ given in the text. We use
the measured value for $\langle S \rangle$ in (\ref{final}) hence there is no additional free parameter.
The dashed line is a constant (guide to the eye). Inset: blow-up of main plot.
 The best fit to a pure power law would give $\tau=1.23$, and to a power law times exponential $\tau=1.2$.}
\label{f:P-s}
\end{center}
\end{figure}%
\begin{table}[b]
\begin{center}
\begin{tabular}{|c|c|c|c|}
\hline
RF & $r_{2}$ & $r_{3}$ & $r_{4}$\\
 \hline
 mean field & 3 & 1.67 & 1.4  \\
 \hline 
 $d=3$, eq.~(\ref{rn})  &2.33 &1.54 &1.34 \\
 \hline
 $d=3$, numerics & $2.25{\pm} .05{\pm}.2$ & $1.48{\pm} .04{\pm}.14$ &$1.27{\pm}.02{\pm}.13$ 
 \\
\hline
 $d=2$,  eq.~(\ref{rn})  &1.66 &1.42 & 1.28 \\
 \hline
 $d=2$, numerics  &$1.95{\pm}.02{\pm}.06$ &$1.38{\pm}.02{\pm}.06$ &$1.21{\pm}.02{\pm}.06$ \\
 \hline\end{tabular}
\end{center} 
\vspace*{-3mm}
\caption{Universal amplitude ratios with statistical and systematic errors (in this order) for numerics; there  is a systematic error since the  measured ratios decrease with decreasing mass. For $d=2$,
the decrease which we take as systematic error  was measured from  masses $m^{2}=0.025$, $m^{2}=0.00125$, and $m^{2}=0.000625$ (whose values are given).
For $d=3$,  the corresponding one  is measured for the two smallest masses $m^{2}=0.0025$ and $m^{2}=0.00125$ (with values from the latter).
}\label{Universal ratios}
\end{table}%
\begin{eqnarray}\label{rn}
r_n:&=& \langle S^{n+1} \rangle \langle S^{n-1} \rangle \langle S^{n} \rangle^{-2} = \frac{2 n-1}{2 n -3} \\
&& - \frac{\epsilon}{3}(1-\zeta_1) \frac{ n \Gamma(n-\frac{3}{2}) + \sqrt{\pi} \Gamma(n-1)}{(2 n - 3)^2
\Gamma(n-\frac{3}{2}) }  + O(\epsilon^2)\ ,\nonumber
\end{eqnarray}
for any real $n>3/2$, with $\zeta_1=1/3$ for RF, $\zeta_1=0$ for RP and $\zeta_1=0.283$ for RB. Upon inversion
of the Laplace transform one finds:
\begin{equation}\label{final}
P(S) = \frac{\left<S\right>}{2 \sqrt{\pi}} S_m^{\tau-2} A S^{-\tau} \exp\!\left(C \sqrt{\frac{S}{S_m}} -
\frac{B}{4} \left[\frac{S}{S_m}\right]^\delta\right)
\end{equation}
for $S \gg S_0$, with $C=- \frac{1}{2} \sqrt{\pi} \alpha$, $B = 1-\alpha(1+\frac{\gamma_{\mathrm{E}}}{4})$, $A = 1+
\frac{1}{8} (2-3 \gamma_{\mathrm{E}} ) \alpha$, $\gamma_{\mathrm{E}}=0.577216$, and  exponents:
\begin{eqnarray}\label{a64}
&& \tau = \frac{3}{2} + \frac{3}{8} \alpha =  \frac{3}{2} - \frac{1}{8} (1 - \zeta_1) \epsilon + O(\epsilon^2)
\\
&& \delta=1 - \frac{\alpha}{4} = 1 + \frac{1}{4} (1 - \zeta_1) \epsilon\ .
\end{eqnarray}
Note that the decay of large avalanches becomes stretched (sub)exponential (in $d=0$ for RF, $\delta=3$). We
note that our result for $\tau$ agrees to $O(\epsilon)$ with the conjecture
\begin{equation}\label{conj}
\tau=2 - \frac{2}{d + \zeta}
\end{equation}
equivalent to  $\rho=2$. It was  presented for {\it depinning} \cite{ZapperiCizeauDurinStanley1998} and for the
$\tau_s=4/3$ exponent of the number of topplings in a sandpile model in $d=3$ \cite{dharLERW}, which may be
compared to CDWs. Since assumptions leading to (\ref{conj}) are not rigorous, our first-principle calculation
confirms this to one loop, and leaves open the possibility of higher-loop corrections. Our results are
straightforwardly extended to the case of LR elasticity \cite{kwpld}.

\begin{figure}[b]
\begin{center}
\fig{0.9\columnwidth}{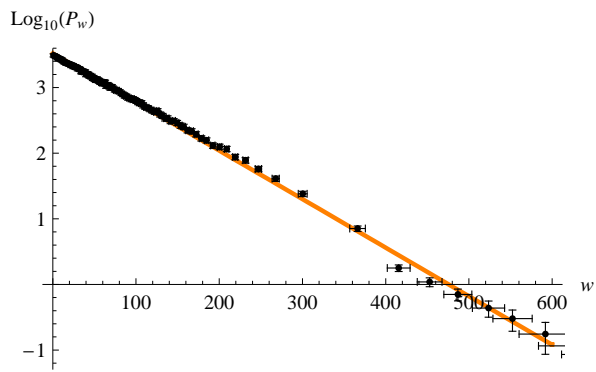} \caption{Distribution of intervals between jumps, in units of the step-size $\delta w=0.002$.
Error-bars are $3\sigma$-errors. (The last 5 boxes are overlapping.)} \label{f:P-w}
\end{center}
\end{figure}
Exact numerical calculation of minimum-energy interfaces has been performed using the discrete models and
algorithms as described in \cite{MiddletonLeDoussalWiese2006} were details can be found. The (corrected)
distribution is presented on figure \ref{f:P-s}, for the example of RF in $d=2$. We measure
\begin{equation}
\tau = 1.25\pm0.02 \  \mbox{(RF, $d=2$)}\ ,  \quad \tau=1.37\pm0.03 \ \mbox{(RF, $d=3$)}
\end{equation}
This is compatible with eq.\ (\ref{conj}).
 Note that the extra
stretched-exponential term  $C$ in (\ref{final}) (which could not be interpreted as summation of a
pre-exponential power) leads to a bump which can clearly be seen in the numerics on figure \ref{f:P-s}. Finally
we have measured (see Fig.~\ref{f:P-w}) the distribution of the intervals between successive jumps (occuring at
positions $w=w^i$) and found it to be very close to a pure exponential.

To conclude, using Functional RG we have performed an expansion around the upper critical dimension to obtain
the avalanche or shock distribution in the statics. It compares well with the numerics. Preliminary results
\cite{us_dynamics} indicate that the above mean-field and 1-loop results also hold for depinning (with the
corresponding values for $\zeta$); 2-loop calculations are in progress to further check the conjecture
(\ref{conj}) and quantify the difference between static and dynamic avalanches.

We acknowledge very useful discussions with Karin Dahmen and Andrei Fedorenko and support from NSF Grants
0109164 and 0219292 (AAM) and ANR program 05-BLAN-0099-01 (PLD and KW).

\end{document}